\documentclass[a4paper,fleqn,usenatbib,useAMS]{mnras}
\input epsf.sty

\usepackage{graphicx}
\usepackage{amsmath,amssymb}
\usepackage[]{natbib}
\usepackage{xcolor}

\title{Limitation of symmetry breaking by gravitational collapse: the revisit of Lin-Mestel-Shu instability}
\author[Tirawut Worrakitpoonpon]{Tirawut Worrakitpoonpon$^{1,2}$\thanks{E-mail: worraki@gmail.com} \\
  $^{1}$School of Physics, Institute of Science, Suranaree University of Technology,
  Nakhon Ratchasima 30000, Thailand \\
  $^{2}$Faculty of Science and Technology, Rajamangala University of
  Technology Suvarnabhumi, Nonthaburi 11000, Thailand}

\pubyear{2020}

\begin{document}
\label{firstpage}
\pagerange{\pageref{firstpage}--\pageref{lastpage}} \pubyear{\the\year}

\maketitle

\begin{abstract}
  We revisit the topic of shape evolution during the spherical
  collapse of an $N$-body system. 
  Our main objective is to investigate the critical particle number
  below which, during a gravitational collapse, the amplification of triaxiality
  from initial fluctuations is effective, and above which it
  is ineffective. To this aim, we develop the Lin-Mestel-Shu theory for a
  system of particles initially with  isotropic velocity dispersion and 
  with a simple power-law density profile.
  We first determine, for an unstable cloud,  two radii corresponding to 
  the balance of  two opposing forces and their fluctuations: such radii 
  fix the sizes of the non-collapsing region and the triaxial seed
  from density fluctuations. We hypothesize that the  triaxial degree 
  of the final state 
  depends on which radius is dominant prior to the collapse phase 
  leading to a different scheme of the self-consistent
  shape evolution of the core and the rest of the system.
  The condition where the two radii are equal therefore identifies
  the critical particle number, which can be expressed as the function of the 
  parameters of initial state.
  In numerical work, we can pinpoint such a critical number
  by comparing the virialized flattening with the initial flattening.
  The difference between these two quantities
  agrees with the theoretical predictions only for the power-law density
  profiles with an exponent in the range $[0,0.25]$. For higher exponents,
  results suggest that the critical number is above the range of simulated $N$.
  We speculate that there is an additional mechanism, related to
  strong density gradients that increases further the flattening,
  requiring higher $N$ to further weaken the initial fluctuations.

\end{abstract}


\maketitle

\section{Introduction} \label{intro}

The gravitational collapse model is widely considered to be a potential
explanation for the origin and early behavior of elliptical
galaxies. The Lin-Mestel-Shu (LMS) theory represents a 
well-established
framework for this model \citep*{lin+mestel+shu_1965}.
The numerical solutions of the self-consistent equations of motion
explain that,  in a collapse starting from rest,
both unstable prolate and unstable oblate ellipsoids
can enhance their flattening.
From a dynamical point of view, the relaxation mechanism for
out-of-equilibrium systems, known as violent relaxation theory,
was provided by \citet{lynden_bell_1967}. It aims, at first, to
resolve the mystery of light distribution of elliptical galaxies,
which appears to conform
with the universal de Vaucouleurs' profile \citep{de_vaucouleurs_1953},
and to address the problem of their formation time-scale,
which appears to be longer than the age of the universe.
This mechanism drives the system
to an unthermalized quasi-stationary state (QSS)
that depends on the initial conditions within a time-scale
comparable to the dynamical time.
Early tests of these hypotheses by \citet{aarseth+binney_1978}, 
in which the initial systems were unstable flat ellipsoids, 
have shown the formation of relaxed structures with a
wide range of ellipticity and other properties. Then, the series of gravitational
collapses with diversified initial conditions have demonstrated
that the final state may have
ellipticity up to E4 \citep{van_albada_1982, mcglynn_1984}.

In the following years, various systematic studies
have allowed us to better understand
the dependences on some initial system parameters such as 
the rotational speed \citep{aguilar+merritt_1990}, 
the virial ratio \citep{theis+spurzem_1999, roy+perez_2004}, 
the power-law index of the density profile
\citep{cannizzo+hollister_1992, boily+athanassoula_2006,
  sylos_labini+benhaiem+joyce_2015}, or the velocity anisotropy
\citep{barnes+lanzel+williams_2009}.
By choosing the velocity dispersion of the initial state as free parameter,
then a compilation of past results suggests that one can roughly
classify the relaxation mechanism as leading to strong or weak
symmetry breaking. Systems that relax starting from a low supporting pressure
tend to strongly break the initial symmetry while systems initially
possessing a high velocity dispersion tend to remain close to this symmetry.
Some studies are able to partition the strong and weak symmetry breaking
regimes using the virial ratio or the velocity anisotropy parameter
\citep{min+choi_1989, aguilar+merritt_1990, boily+athanassoula_2006,
  barnes+lanzel+williams_2009}.
In addition to the difference of the flattening amplification, the relaxed
density profile and velocity anisotropy are also found to be different
\citep{mcglynn_1984, aguilar+merritt_1990, sylos_labini_2012, sylos_labini_2013}.

While the dependences on initial state parameters are commonly studied,
effects from finite-$N$ fluctuations that are
embedded in the initial conditions have not attracted much attention,
although several authors have noticed their role. 
For instance, this issue has been raised by \citet{aarseth+lin+papaloizou_1988}
and then it was revisited by \citet{boily+athanassoula+kroupa_2002}
and \citet{joyce+marcos+sylos_labini_2009}.
In particular in this latter work, it was shown that
finite-$N$  fluctuations have a key role in the system  
 evolution determining the  post-collapse ejection of
matter and energy from the system. 
The effect of Poissonian fluctuations
on the final system's shape has been studied in the works of
\citet{roy+perez_2004} and \citet{boily+athanassoula_2006},
where it was concluded to have a negligible effect provided that
the number of system particles $N$ is high enough. That conclusion was revised
by \citet{worrakitpoonpon_2015, benhaiem+sylos_labini_2015}
where it was reported, by making accurate systematic tests, that
the flattening and other parameters of the virialized states constantly vary
in a wide range of $N$. In addition, \citet{benhaiem_et_al_2018} 
have shown that by reducing the density fluctuations (i.e. increasing $N$),
the onset of the radial orbit instability can be turned off.
This result emphasizes the significance of statistical 
density fluctuations in the   violent relaxation mechanism.

In this work, we revisit the spherical collapse paradigm, focusing
on the morphological evolution of an isolated self-gravitating system of
particles toward an ellipsoidal stationary state.
Our main purpose is to find the critical particle number
separating between effective and ineffective evolution to triaxial 
shape of an isolated system. We study this problem by making use of  
an extensive set of $N$-body simulations. Comparing to past studies
with a similar purpose, we have enlarged the analysis in several ways.
Firstly, we develop the collapse model of an $N$-particle cloud
based on the inside-out evolution scenario of the LMS theory 
by considering the role of fluctuations both in the density and
velocity spaces. Secondly, we propose a reliable
measurement to differentiate the numerical 
results that  allows us  to test our hypothesis.

The article is organized as follows. In Section \ref{collapse_eq},
we describe the initial conditions that we have considered for the
numerical evolution. In addition, we present the theoretical model that we
develop from the LMS framework. Our analysis finally leads to identifying
the critical particle number as a function of system parameters:  
such an analysis will be then compared with the results of numerical simulations.
The technical details of the simulations including the preparation and
accuracy control are given in Section \ref{nume_para}.
In Section \ref{nume}, we report the numerical results of the shape evolution
by violent relaxation and compare them with the theoretical predictions.
Finally in Section \ref{conclusion} and \ref{discussion},
we provide respectively the conclusions of our study and some further
discussion about how our results relate to past work.

\section{Analysis of unstable particle clouds}
\label{collapse_eq}

\subsection{Properties of initial conditions} 
\label{analysis_ic}

Initial conditions are represented by  spherical clouds of particles with a 
power-law density profile with a cut-off of the type
\begin{equation}
  \rho(r)=\frac{(3-\alpha)M}{4\pi r_{0}^{3-\alpha}r^{\alpha}} \label{rho_r}
\end{equation}
where $M$ is the total mass, $0 \le \alpha \le 3$ is the power-law index
and $r_{0}$ is the cut-off radius: the case $\alpha=0$ corresponds to
the uniform case. From Eq. (\ref{rho_r}), 
we derive integrated mass in a sphere of radius $r$
\begin{equation}
  M(r)=\frac{r^{3-\alpha}}{r_{0}^{3-\alpha}}M \label{m_r} 
\end{equation}
and the average density profile
\begin{equation}
  \bar{\rho} (r)\equiv \frac{M(r)}{\frac{4}{3}\pi r^{3}}
  =\frac{3}{3-\alpha}\rho (r).
  \label{rho_bar_r}
\end{equation}
The binding potential energy of the system is then
\begin{equation}
  U_{0}=-\int_{0}^{r_{0}}\frac{GM(r)\rho (r)(4\pi r^{2}dr)}{r}
  =-\frac{(3-\alpha)GM^{2}}{(5-2\alpha)r_{0}}. \label{u_0}
\end{equation}
The systems have initially  isotropic top-hat
velocity distribution that can be written as
\begin{equation}
  \theta (v)=\frac{3M}{4\pi v_{0}^{3}} \label{theta_v} \;\; \mbox{for $v \le v_{0}$} 
\end{equation}
where $v_0$ is the cut-off velocity. The normalization condition
is then $M=\int_{0}^{v_{0}}\theta (v) (4\pi v^{2}dv)$. 
The corresponding kinetic energy can be calculated to be
\begin{equation}
  T_{0}=\frac{1}{2}\int_{0}^{v_{0}}v^{2}\theta (v)(4\pi v^{2}dv)
  =\frac{3}{10}Mv_{0}^{2}. \label{t_0}
\end{equation}
The uniform velocity distribution $\theta (v)$ has dispersion $\sigma$
such that
\begin{equation}
  \sigma^{2}=\frac{1}{M}\int_{0}^{v_{0}}v^{2}\theta (v)(4\pi v^{2}dv)
  =\frac{3}{5}v_{0}^{2}. \label{sigma_v}
\end{equation}

The pressure due to random motions in a spherical system
of radius $r$ is
\begin{equation}
  p(r)=\frac{1}{3}\sigma^{2}\rho(r). \label{p_r}
\end{equation}
Note that for $\alpha=0$, the pressure is uniform as the density.
From Eq. (\ref{u_0}) and (\ref{t_0}), we find that initial virial ratio reads 
\begin{equation}
  b_{0}\equiv \frac{2T_{0}}{|U_{0}|}=\frac{(5-2\alpha)r_{0}\sigma^{2}}{(3-\alpha)GM}.
  \label{b_0}  
\end{equation}

\subsection{Equilibrium of forces and effects of density fluctuations}
\label{collapse_alphanonzero}

Let us firstly consider spherical clouds with $\alpha>0$. 
The inward gravitational force per unit mass at any radius is
\begin{equation}
  f_{g}(r)=\frac{GM(r)}{r^{2}}=\frac{GMr^{1-\alpha}}{r_{0}^{3-\alpha}}.
  \label{f_g_non0}
\end{equation}
The outward pressure force per unit mass can also be calculated
using Eq. (\ref{p_r}) 
\begin{equation}
  f_{p}(r)=-\frac{1}{\rho (r)}\frac{dp}{dr}=\frac{\alpha \sigma^{2}}{3r}.
  \label{f_p_non0}
\end{equation}
Note that $ f_{p}$ diverges as $r\rightarrow 0$ and decays as $r^{-1}$
for large distance, while  $f_{g}$ is an increasing function of $r$ for $\alpha <1$.

Using  an approach similar to the Jeans' instability, we define the
Jeans-like radius, $r_{J}$, where these two forces are equal,
i.e. $f_{g}(r_{J})=f_{p}(r_{J})$; we obtain
\begin{equation}
  r_{J}=\bigg[ \frac{\alpha (3-\alpha)}{3(5-2\alpha)}\bigg]^{\frac{1}{2-\alpha}}
  b_{0}^{\frac{1}{2-\alpha}}r_{0}.   \label{r_h_nonzero}
\end{equation}
This radius separates the non-collapsing, i.e. for  $r<r_{J}$, 
and the gravitationally unstable, i.e.  $r> r_{J}$, regions. 
We note that $r_{J}$
does not depend on the particle number but it increases with $b_{0}$. 

Next, let us consider the role of 
discrete fluctuations in a system of $N$  equal mass particles.
If the particles are randomly placed, density 
fluctuations  are Poissonian with a relative amplitude
\begin{equation}
  \frac{\delta \bar{\rho} (r)}{\bar{\rho} (r)}\sim
  \frac{\delta \rho (r)}{\rho (r)}\sim \frac{\delta M(r)}{M(r)}
  \sim \frac{1}{\sqrt{N(r)}} \label{estim_fluc}
\end{equation}
where $N(r)$ is the particle number enclosed inside $r$. Using the 
integrated mass profile (\ref{m_r}), we estimate $N(r)$ to be
\begin{equation}
  N(r) = \frac{r^{3-\alpha}}{r_{0}^{3-\alpha}}N    \label{n_r}
\end{equation}
where $N$ is the total particle number. 
Fluctuations of $f_{g}$ at any radius are given by
\begin{equation}
  \delta f_{g}(r)=\frac{G\delta M(r)}{r^{2}}. \label{def_df_g}
\end{equation}
Following the approximation (\ref{estim_fluc}), we obtain $\delta f_{g}$ in 
terms of system parameters as
\begin{equation}
  \delta f_{g}(r)\sim \frac{GM}{r_{0}^{(3-\alpha)/2}r^{(1+\alpha)/2}N^{1/2}}
  \label{df_g_r}
\end{equation}
that explicitly depends on $N$. Note that
$\delta f_{g} \rightarrow 0 $ for $N\rightarrow \infty$ as in this limit,
density fluctuations vanish. Eq. (\ref{df_g_r}) is also useful to quantify 
the deviation from spherical symmetry, which then becomes the source (or seed)
of the amplification of structure triaxiality.

In the same approximations, we calculate the fluctuations of
outward force $f_{p}$ by 
\begin{equation}
  \delta f_{p}(r)=\frac{\alpha}{3r}\delta\sigma^{2} \label{def_f_p_nonzero}
\end{equation}
which involves the fluctuations of $\sigma^{2}$. To estimate it, we recall the
definition of the velocity dispersion of an $N$-particle system
and write it in terms of the expected value and its deviation as 
\begin{equation}
  \sum_{i,j=1}^{j=N}\frac{(v_{i,j}-\bar{v}_{i})^{2}}{N}=\sum_{i}\overline{v_{i}^{2}}
  -\sum_{i}\bar{v}_{i}^{2}=\overline{v^{2}}-\bar{v}^{2}
  \equiv \sigma^{2}-\delta\sigma^{2} \label{estim_variance}
\end{equation}
where $i$ is summed over all three-dimensional velocity components of
particles and $j$ is the index of particles. For the isotropic velocity distribution,
$\sigma^{2}$ converges to $\overline{v^{2}}$ as $N\rightarrow\infty$. Thus
\begin{equation}
  \delta\sigma^{2}\sim \bar{v}^{2} \label{delta_sigma_estim}
\end{equation}
where $\bar{v}$ can be estimated by the standard deviation of the mean.
For the population of particles inside $r$, we obtain the scaling
of $\delta\sigma^{2}$ via $N(r)$ as
\begin{equation}
  \frac{\delta \sigma^{2}}{\sigma^{2}}\sim\frac{1}{N(r)}. \label{estim_vbar}
\end{equation}
Using the expression of $N(r)$ given in Eq. (\ref{n_r}), the 
fluctuations of $f_{p}$ induced by finite-$N$ effects reads
\begin{equation}
  \delta f_{p}(r)\sim \frac{\alpha r_{0}^{3-\alpha}\sigma^{2}}{3r^{4-\alpha}N}.
  \label{df_p_r}
\end{equation}
Thus  $\delta f_{p}$ decays more rapidly with $N$ than $\delta f_{g}$ 
(see Eq. (\ref{df_g_r})). We note the same counter-action between
$\delta f_{g}$ and $\delta f_{p}$ as with that between $f_{g}$ and $f_{p}$:
$\delta f_{p}$ dominates close to the center while at large radius, $\delta f_{g}$
becomes dominant. We then define another radius, $\tilde{r}$,
where $\delta f_{g}(\tilde{r})\sim \delta f_{p}(\tilde{r})$ and obtain
\begin{equation}
  \tilde{r}\sim \bigg[ \frac{\alpha (3-\alpha)}{3(5-2\alpha)} \bigg]^{2/(7-3\alpha)}
  \frac{b_{0}^{2/(7-3\alpha)}}{N^{1/(7-3\alpha)}}r_{0}. \label{r_instab}
\end{equation}
This radius is, unlike $r_{J}$, shifting inwards as $N$ increases for $\alpha$
below $7/3$. It indicates another border above which the asymmetry seeded by
the density fluctuations is dominant over the pressure fluctuations.
We regard the former fluctuations as the seed of triaxiality with 
effective size $\tilde{r}$.
The development of triaxiality of initial sphere
is then proposed as described by the following scenario.
While the sphere is collapsing, the boost of triaxiality deep inside
the system, i.e. the core, around $\tilde{r}$ results in the alteration of
the central gravitational field to become more ellipsoidal, which then
enhances the eccentricity of the entire mass accordingly while it is
falling to the center. For this scenario to occur, we hypothesize
that $\tilde{r}\gtrsim r_{J}$ so that the development of triaxiality proceeds
effectively starting from $\tilde{r}$. Otherwise,
if $\tilde{r} \lesssim r_{J}$, the triaxial seed is shielded by a stable
radius marked by $r_{J}$ so the core eccentricity is not effectively
developed. This makes the entire structure weakly (or not at all)
amplified due to the lack of a strong asymmetric field.
Thus, the condition where $\tilde{r}\sim r_{J}$ yields
the transition point separating between effective and ineffective core
symmetry breaking, which plays a crucial role to determine the 
  morphological evolution scheme of the entire structure.
By that condition, we express the transition point in empirical terms of
the critical particle number, $N_{c}$, indicating the number
above which the density fluctuations cease to be developed as
\begin{equation}
  N_{c}(b_{0};\alpha)\sim \Big[
    \frac{3(5-2\alpha)}{\alpha (3-\alpha)}\Big]^{\frac{3-\alpha}{2-\alpha}}
  b_{0}^{-\frac{3-\alpha}{2-\alpha}}.   \label{n_c_nonzero}
\end{equation}
It becomes clear from Eq. (\ref{n_c_nonzero}) that $N_{c}$ decreases
with $b_{0}$. In other words, this number separates between
gravity fluctuation- and pressure-dominated core collapse that leads to
different fates of the system configuration.
We note that by this expression, $N$ diverges as $\alpha =0$. 
This particular case will be handled in Section \ref{collapse_alphazero} with
additional suppositions.

\subsection{The case of an initially uniform spherical mass distribution}
\label{collapse_alphazero}

We consider the same mechanism as described in Section \ref{collapse_alphanonzero}
for uniform density. In principle,
the lack of a density gradient leads to the absence of a
pressure force (see Eq. (\ref{f_p_non0})). However, the finite $N$
gives rise to the density fluctuations, making the effective density become
$\rho (r) = \rho_{0}+\delta\rho(r)$ where $\delta\rho (r)$
can be estimated by Eq. (\ref{estim_fluc}).
The pressure force from the density fluctuations corresponds to
\begin{equation}
  f_{p}(r)\sim -\frac{\sigma^{2}}{3\rho_{0}}\frac{d\delta \rho}{dr}
  = \frac{\sigma^{2}r_{0}^{3/2}}{2r^{5/2}N^{1/2}}. \label{f_p_zero}
\end{equation}
This force is now the decreasing function of $N$.
We consider this pressure force as the dominant one because the
zeroth-order force is missing due to the uniform density.
We then obtain $r_{J}$ for $\alpha =0$ to be
\begin{equation}
  r_{J}\sim \bigg(\frac{3}{10}\bigg)^{2/7} \frac{b_{0}^{2/7}}{N^{1/7}}r_{0}
  \label{r_h_zero}
\end{equation}
which is now decreasing with $N$. The fluctuations of $f_{p}$
  can be obtained via $\delta\sigma^{2}$ as
\begin{equation}
  \delta f_{p}\sim \frac{r_{0}^{3/2}}{2r^{5/2}N^{1/2}}\delta\sigma^{2}
  \sim \frac{\sigma^{2}r_{0}^{9/2}}{N^{3/2}r^{11/2}} \label{df_p_zero}
\end{equation}
which decays with $N$ more rapidly than $f_{p}$.
We then have $\tilde{r}$ by comparing Eq. (\ref{df_g_r}) for $\alpha =0$
with Eq. (\ref{df_p_zero}) to be
\begin{equation}
  \tilde{r}\sim \bigg( \frac{3}{10}\bigg)^{1/5}\frac{b_{0}^{1/5}}{N^{1/5}}r_{0}.
  \label{r_instab_zero}
\end{equation}
From the obtained $r_{J}$ and $\tilde{r}$, we observe the same
counter-action similar to the case with $\alpha >0$:
$r_{J}$ and $\tilde{r}$ conceal the other at large and small $N$,
respectively. By the condition where $r_{J}=\tilde{r}$,
  we obtain the critical particle number for $\alpha=0$ as
\begin{equation}
  N_{c}(b_{0};0)\sim \Big( \frac{10}{3}\Big)^{\frac{3}{2}}b_{0}^{-\frac{3}{2}}.
  \label{n_c_zero}
\end{equation}
This critical number also decreases with $b_{0}$
as with $\alpha >0$.
We remark that the same exponent of $b_{0}$ can also be retrieved
from $N_{c}(b_{0};\alpha)$ in Eq. (\ref{n_c_nonzero})
if one puts $\alpha\rightarrow 0$. However, the discrepancy 
arises in the pre-factor which makes $N_{c}$ diverge.
This is because, without the density fluctuations, the same
treatment from Section \ref{collapse_alphanonzero} gives both
$f_{p}$ and $\delta f_{p}$ equal to zero.

\subsection{Extended scope from the LMS theory}
\label{relevance_lms}

Before we bring Section \ref{collapse_eq}
to an end, we will discuss in detail how our hypothesis
conceptually follows and extends the original scope
of \citet*{lin+mestel+shu_1965}. In their work, 
the self-consistent equations of motion are first parametrized
by the axis ratio and then the subsequent evolution is traced.
Given any amount of initial flattening, the axis ratio can be greatly
amplified in the course of a gravitational collapse governed by
a self-consistent asymmetric field. An evolutionary track of
the axis ratio also depends on its starting value.

While the original LMS scope trivially imposes the initial flattening
as a seed of further evolution, we consider here the Poissonian density
fluctuations as the initial seed with effective flattening that can
be scaled by $N$. In the absence of the velocity dispersion, various
numerical experiments have verified the influence of $N$ on final triaxiality
\citep{worrakitpoonpon_2015, benhaiem+sylos_labini_2015}, as directly
implied by the LMS theory.
In this work, we introduce the velocity dispersion to the initial state
and consider both its primary mechanical effect (i.e. the pressure)
and its effect associated to the Poissonian noise.
Finding the points
where the forces and their fluctuations are balanced yields two
characteristic radii that fix the sizes of the non-collapsing region
and the unstable triaxial seed by density fluctuations.
While the meaning of the former radius can be retrieved from the basis of
Jeans' instability, the meaning of the latter one is proposed anew.

Knowing these two radii that have the different roles in the
initial states, we hypothesize that the triaxiality of final states
depends on which radius is dominant prior to the collapse phase.
Thus, the transition point between two regimes of shape evolution
corresponds to the point where they are equal.
Our criterion is parametrized by
the critical particle number as a function of system parameters.
Note that our analysis does not predict the detailed evolution from 
initial state to final state as demonstrated by
\citet*{lin+mestel+shu_1965} where the shape parameter is precisely
tracked from the initial point to the maximum collapse.
We propose that the shape evolutions by that mechanism proceed
differently above and below the critical particle number.
This hypothesis will be later tested by numerical simulations with
suitable numerical parameters (see Section \ref{nume}).

The point that requires proper clarification is that the
introduced velocity dispersion does not dismiss the initial
gravitational instability, which takes a major role during the
development of triaxiality. Our system always starts from
a sub-virialized state. Starting with this instability, 
we propose two scenarios of the self-consistent shape evolutions
that follow. An alternate way where the enhancement of ellipticity by
the collapse can be inactivated by sufficiently high velocity dispersion
can be considered as an enlargement of the original LMS scope.

\section{Simulation set-up} \label{nume_para}
  
\subsection{Initial conditions and units} \label{sec_ic}

Isolated systems of self-gravitating $N$ particles following
the spherically symmetric density profile (\ref{rho_r}) and the
velocity distribution (\ref{theta_v}) are constructed as initial conditions
for our study. Three-dimensional position and velocity are assigned
randomly to each particle following those distributions
so that the Poissonian noise in both spaces is preserved.
Initial velocity dispersion is controlled by the initial virial ratio $b_{0}$,
which is numerically adjusted in accordance with the given random configuration.
We inspect the cases with $\alpha \in [0,0.5]$ in order
to explore two different conditions proposed in
Section \ref{collapse_alphanonzero} and \ref{collapse_alphazero}. 

Length and mass scales are chosen so that the initial radius of sphere $r_{0}$
and the initial mean density $\bar{\rho}_{0}$, that is calculated beneath $r_{0}$,
are $0.5$ and $1$, respectively, for any $\alpha$. The time unit is specified to be
\begin{equation}
  t_{d}=\sqrt{\frac{3\pi}{32G\bar{\rho}_{0}}} \label{t_dyn}
\end{equation}
where $G$ is the Newtonian gravitational constant. Note that this
time scale corresponds to the free-fall time for a cold uniform sphere.

\subsection{Numerical integration and accuracy control} \label{para_nume}
Equations of motion of particles are integrated by GADGET-2 in public version
(see \citealp{springel+yoshida+white_2001, springel_2005}). An isolated sphere
of particles is placed in open space without any expansion of its background.
In the code, the Newtonian force is spline-softened below a pre-defined softening
length $\varepsilon$ so that, below this length, the force converges to
zero at zero separation. We choose this length to decrease with $N$ as
\begin{equation}
  \varepsilon=0.0028N^{-1/3} \label{soften_length}
\end{equation}
so that $\varepsilon$ keeps up with initial inter-particle distance which
can approximately be scaled by the same factor of $N$. The small numerical
pre-factor before $N$ is included to assure that $\varepsilon$ is well below the
minimum inter-particle distance attained during the maximum contraction.
We use a small opening angle so that the force is as close as possible
to that calculated by direct summation. The time-step before $1.43 \ t_{d}$
is controlled to be below $t_{d}/35000$ since this is
the period of strong collapse. The limitations of opening-angle
and time-step during this stage
are important because the development of triaxiality relies on a delicate
anisotropic force as suggested by our hypothesis. Afterwards, the
time-step is extended but it is never longer than $t_{d}/8000$. With those 
integration parameters, total energy is conserved within $0.1\%$ of
deviation throughout the simulation. We remark that the deviation is
highest, though still within the indicated level of precision, when $\alpha =0$
and $b_{0}$ is low. Away from that limit, the deviation is typically not more
than $0.02\%$. This is actually the anticipated outcome since the collapse of
the cold uniform sphere approaches the singularity where the forces are strong
and vary rapidly.

\section{Numerical results} \label{nume}

In general, the amplitude of the statistical fluctuations in an $N$-body system
increases when $N$ decreases. One way to overcome
the small-$N$ fluctuations that may arise in the result is to perform a higher
number of realizations for lower $N$. If necessary, the time average
could be taken in addition to the ensemble average.
Typical numbers of realizations for systems with
$N=1000$, $2000$, $4000$, $8000$, $16000$, $32000$, $64000$ and $128000$ are
$70$, $40$, $30$, $30$, $6$, $4$, $2$ and $1$, respectively.

\subsection{Virialization of spherical collapse} \label{collapse_general}

\begin{figure}
  \begin{center}
    \includegraphics[width=8.5cm]{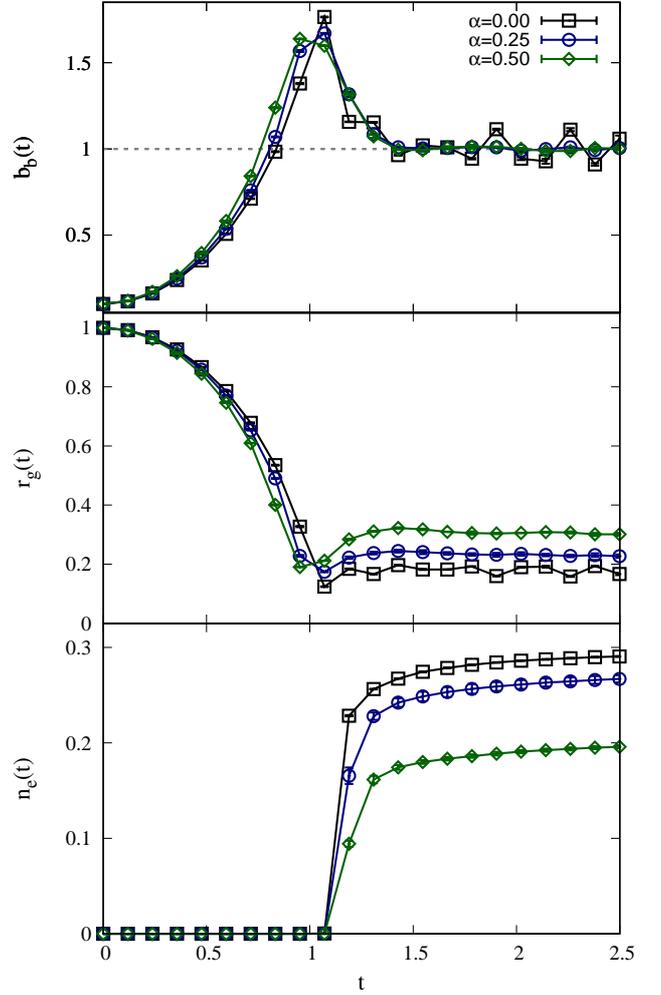}
  \end{center}
  \caption{Ensemble-averaged temporal evolutions of the virial ratio of
    bound mass $b_{b}$ (top panel), the gravitational radius of bound mass $r_{g}$
    (middle panel) and the fraction of ejected mass $n_{e}$ (bottom panel) for 
    different indicated $\alpha$ with $b_{0}=0.1$ and $N=64000$.
    Size of error bars corresponds to the standard deviation of the mean.}
  \label{fig_3parameters}
\end{figure}

We first consider the overall process of violent relaxation for different
$\alpha$ and $b_{0}$. Shown in Fig. \ref{fig_3parameters} are ensemble-averaged
temporal evolutions of the virial ratios of bound particles $b_{b}$,
the gravitational radii of bound particles $r_{g}$ and the
fractions of ejected particles $n_{e}$ around the violent relaxation for different
$\alpha$ with $b_{0}=0.1$ and $N=64000$. The definition of $r_{g}$ is given by
\begin{equation}
  r_{g}=\frac{GM_{b}^{2}}{|U_{b}|} \label{rg_def}
\end{equation}
where $M_{b}$ and $U_{b}$ are the mass and potential energy computed from bound particles.
This radius is presented as the unit of distance in the initial value.
First, we see that the virializations of the three cases are similar in the way
that $b_{b}$ increases from $b_{0}$ by collapse and goes beyond $1$
at $t\sim 1$. It then relaxes down to a virialized state around $1.5 \ t_{d}$.
Increasing $\alpha$ makes the increment slightly faster and $b_{b}$ reaches
slightly lower maximum value. The evolution of $r_{g}$ is consistent with
that of $b_{b}$. From its initial value,
it decreases and attains a minimum at the time when
$b_{b}$ reaches its maximum. It is then brought up to its virialized value.
We remark that both the minimum attained and final $r_{g}$ tend to be
lower if $\alpha$ decreases, which indicates that the collapse and resulting
central core are more compact as $\alpha\rightarrow 0$.
This can be explained in the following way. 
In the fluid limit with $b_{0}=0$, we can prove
that all mass is falling to the center in the global free-fall time.
In a point system with $b_{0}>0$, the effects from Poissonian noise
and velocity dispersion cause the free-fall times of particles to disperse,
limiting the compactness from progressing close to a singularity.
Furthermore, if $\alpha>0$, the individual
free-fall time is longer at larger radial distance, so the time
spread is even wider because of $\alpha$ than it would be from Poissonian
noise alone. This factor makes the infall more prolonged,
resulting in the observed variation of $r_{g}$.
The degree of violent collapse also manifests itself in the final values of
$n_{e}$ where the particles are ejected more
as $\alpha$ decreases from $0.5$ to $0$. The ejection mechanism,
described by \citet{joyce+marcos+sylos_labini_2009}, is that the
late-arriving particles gain a gravitational slingshot effect from the
time-varying central potential formed by
the particles that arrive earlier. Thus, the variation
of $n_{e}$ also reflects the concentration formed by different $\alpha$.
This $\alpha$-dependence of collapse properties has been
reported before by \citet{sylos_labini+benhaiem+joyce_2015} for $b_{0}=0$ but
here we find that the similar behavior is maintained even if $b_{0}>0$.

\begin{figure}
  \begin{center}
    \includegraphics[width=8.5cm]{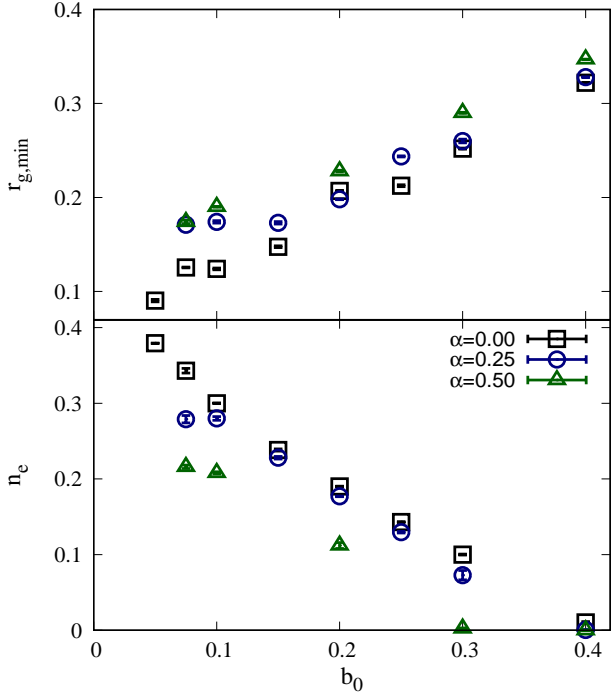}
  \end{center}
  \caption{Ensemble-averaged minimum gravitational radius
    in the collapse $r_{g,min}$    
    (top panel) and $n_{e}$ at $9.52\ t_{d}$ (bottom panel) as function
    of $b_{0}$ for $N=64000$. Size of error bars correspond to
    the standard error of the mean.} \label{fig_collapse_b0}
\end{figure}

Next, we consider the consequence of the violent relaxation from
different $b_{0}$.
Shown in Fig. \ref{fig_collapse_b0} are ensemble-averages of the minimum
gravitational radius in the collapse ($r_{g,min}$) and
$n_{e}$ measured at $9.52 \ t_{d}$,
plotted as functions of $b_{0}$ for different $\alpha$.
All cases include $64,000$ particles.
From the overall viewpoint, the behaviors of both parameters indicate that
increasing $b_{0}$ can moderate the collapse and the ejection.
When $b_{0}\geq 0.2$, the calculated values between $\alpha=0$ and
$0.25$ become closer to each other. This implies that the
influence of mild $\alpha$ is overcome by the initial pressure.
The variation of $n_{e}$ also suggests
that there should be no ejection beyond $b_{0}=0.4$.
With a closer perspective, we remark that the variations are indeed different
at high and low $b_{0}$. When $b_{0}\geq 0.2$, both $r_{g,min}$ and $n_{e}$
vary in the way that we expect from increasing $b_{0}$ since it
provides a stronger outward force against the contraction.
For $b_{0}< 0.2$, we remark that the plots from non-zero $\alpha$
appear to converge to certain values, which are seen more clearly
from $n_{e}$ plots. This result can be explained if we consider that
in the limit where the influence of $b_{0}$ is not significant,
the evolutions of cloud size and concentration mainly depend
on the details of density profile. This leads to the
ordered infall of each mass shell from a different position.
Thus, we generally expect that if $b_{0}\rightarrow 0$, the parameters
should converge to non-zero values that depend strongly on $\alpha$,
as suggested by \citet{sylos_labini+benhaiem+joyce_2015}.
For $\alpha=0$, there is, on the contrary, no sign of convergence 
as both parameters change constantly until the lowest shown $b_{0}$.
The dissimilarity with previous results may possibly be related to
the absence of a density gradient to control the minimum size.
As we approach the cold limit, the evolution to maximum contraction
is regulated by finite-$N$ fluctuations. However,
as suggested by \citet{joyce+marcos+sylos_labini_2009}, it is possible
that there also exists the terminal $r_{g,min}$ and $n_{e}$ 
that now depend on $N$. That value of $b_{0}$ at which they 
converge might be far below those for $\alpha >0$ 
since the effect of density fluctuations is much weaker than that of
the density gradient. Note that the measurement of $r_{g,min}$
can be done more precisely by increasing time resolution of the
snapshot since the attainment to the minimum is momentary in time
unlike $n_{e}$, which is seen to be changing less once it
is virialized. This issue is peripheral to our study.

\subsection{Development of triaxiality for different $\alpha$}
\label{collapse_iota}

In this section, we inspect the development of triaxiality for different
initial conditions. Before we start, we recall that the violent relaxation
of a collapsing cloud generally yields some amount of mass ejection.
We will thus consider the configuration of 
bound particles only. To quantify it, we define 
\begin{equation}
  \iota_{\chi}=\frac{\Lambda_{3,\chi}}{\Lambda_{1,\chi}}-1 \label{iota_def}
\end{equation}
to be the flattening of $\chi$ percent of most bound mass. In that expression,
$\Lambda_{3,\chi}$ and $\Lambda_{1,\chi}$ correspond to the highest and lowest
eigenvalues, respectively, of the moment of inertia tensor calculated from
$\chi$ percent of most bound particles around their respective center of mass.
The flattening of the relaxed structure calculated from $\chi =80$ has been found
suitable since it excludes an amount of loosely bound particles that disrupt the
calculation of the moment of inertia tensor from the majority of particles
\citep{boily+athanassoula_2006, barnes+lanzel+williams_2009, worrakitpoonpon_2015}.
However, the tracking of $\iota_{\chi}$ with different $\chi$ values might still be
useful as it allows us to observe the development of triaxiality at different levels.

\begin{figure}
  \begin{center}
    \includegraphics[width=8.5cm]{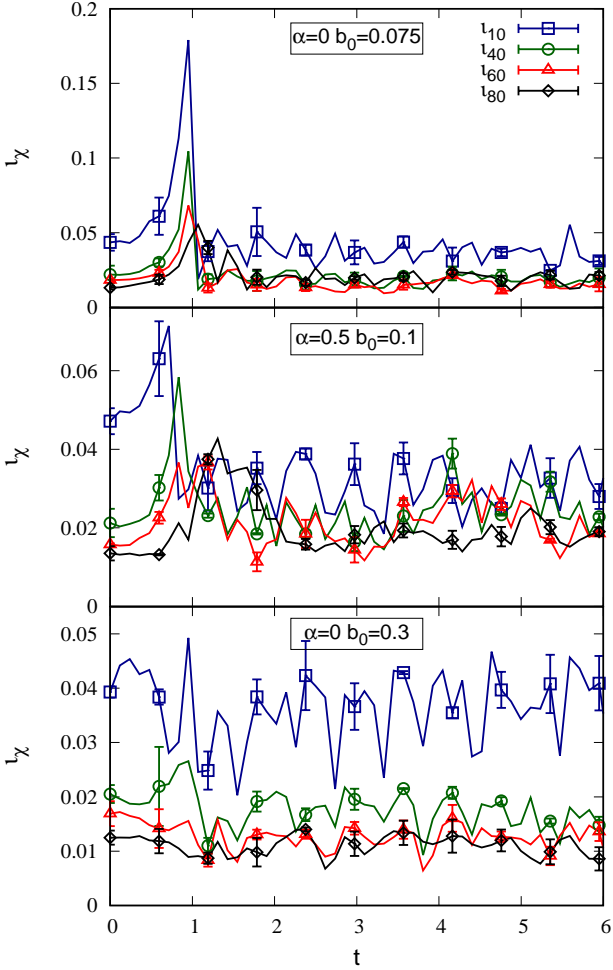}
  \end{center}
  \caption{Temporal evolutions of ensemble-averaged different $\iota_{\chi}$ 
    for $\alpha=0$ and $b_{0}=0.075$ (top panel); $\alpha=0.5$ and
    $b_{0}=0.1$ (middle panel); and $\alpha=0$ and $b_{0}=0.3$ (bottom panel). 
    All of these cases are with $N=64,000$. Size of error bars corresponds to
    the standard deviation of the mean.}
  \label{fig_iota_xx_lowb0}
\end{figure}

To verify this process, temporal evolutions of ensemble-averages of 
$\iota_{\chi}$ for three indicated cases with $64,000$ particles
are shown in Fig. \ref{fig_iota_xx_lowb0}. For $b_{0}\leq 0.1$, $\iota_{\chi}$
is amplified in all levels by anisotropic collapse and reaches the
maximum that is a few times larger than its starting value. It then relaxes
down to a quasi-stationary state afterwards. Fluctuations are still present after
the violent relaxation but the average flattening as evaluated from $\iota_{80}$ is
clearly above its initial one. A major difference between the two cases
is the time $\iota_{\chi}$ takes to attain maximum during the violent relaxation.
For the uniform case, all $\iota_{\chi}$ evolve accordingly and reach their peaks
almost at the same time. For $\alpha=0.5$, the peaks shift forward in time
as a larger fraction is involved.
The different evolutionary patterns of $\iota_{\chi}$ occur because
the local free-fall time is uniform and increasing with radius for $\alpha=0$
and $0.5$, respectively. Thus, for the latter case, higher $\chi$
tends to have longer effective collapse time during
which the triaxiality is amplified.
An analogous plot is illustrated by \citet{boily+athanassoula_2006}
for $b_{0}=0$ and $\alpha=0.5$ that leads to a similar statement as with 
our $\alpha >0$ case. Here we illustrate clearly the difference
between the cases with $\alpha =0$ and $\alpha =0.5$. 
Otherwise, the plot for $b_{0}=0.3$ demonstrates 
weak or absent amplification of $\iota_{\chi}$ during the violent relaxation
before it relaxes to the virialized state in a similar way.
Following our proposed mechanism, the weaker asymmetric field
inside, indicated by $\iota_{10}$, results in the suppressed development
of $\iota_{\chi}$ for larger mass fractions during this early stage.

According to our proposal,
the situations for $b_{0}\leq 0.1$ strictly conform with the original
LMS scenario as we observe the effective inside-out amplification of
$\iota_{\chi}$. With $b_{0}=0.3$, the core triaxiality cannot be
developed much and this consequently limits the amplification of
$\iota_{80}$. That the latter evolution progresses in the
suppressed way affirms our hypothesis that the LMS deformation
process can be inactivated by sufficiently high $b_{0}$,
as discussed in Section \ref{relevance_lms}.

\subsection{Critical particle number in different cases} \label{transition_collapse}

\begin{figure}
  \begin{center}
    \includegraphics[width=8.5cm]{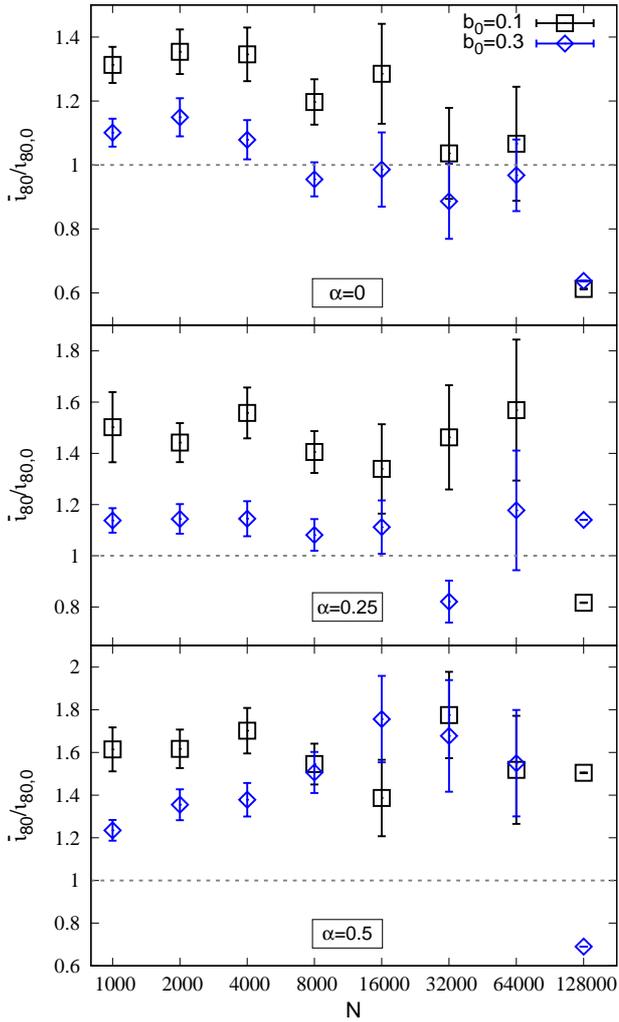}
  \end{center}
  \caption{Ratios between virialized $\iota_{80}$ and initial $\iota_{80}$
    (i.e. $\overline{\iota_{80}}/\iota_{80,0}$) as functions of $N$ for
    different indicated $\alpha$ and $b_{0}$ (see calculation detail 
    in text). Size of error bars corresponds to the standard deviation
    of the mean.}
  \label{fig_ratio_iota_alpha}
\end{figure}

We will verify in this section the existence and the variation
of the critical particle number following
our proposition in Section \ref{collapse_eq}. From the computed $\iota_{80}$ 
in previous section, we choose to evaluate the
amplification effectiveness by comparing this parameter at the stationary state
with the initial one. Given the ensemble-averaged temporal profile
of $\iota_{80}$ in each case, the ratio between
virialized $\iota_{80}$, averaged from
$7.14$ to $9.52 \ t_{d}$ over 20 time-slices, and the initial $\iota_{80}$
(designated by $\overline{\iota_{80}}/\iota_{80,0}$) as function of $N$
for different $\alpha$ and $b_{0}$ is plotted in
Fig. \ref{fig_ratio_iota_alpha}. Inspecting the result in this way allows
us to monitor the intrinsic efficiency more accurately
as this ratio is adjusted by the initial $\iota_{80}$
that has been found to decrease as rapidly as
$N^{-1/2}$ \citep{benhaiem+sylos_labini_2015}. We additionally apply the
time average over the period of the stationary state in order to smooth out
the temporal fluctuations which is seen to be of order final $\iota_{80}$.
Error bars are estimated from the time-averaged standard deviation of the mean
in the same time window. The horizontal lines indicate when initial and
averaged-final $\iota_{80}$ are equal. From a glance at the plot,
we see that a higher $b_{0}$ can suppress more effectively the development
of triaxiality, now parametrized by $\overline{\iota_{80}}/\iota_{80,0}$,
which is seen in all panels and is in line with the result in Section \ref{collapse_iota}.

Considering its variation with $N$, $\overline{\iota_{80}}/\iota_{80,0}$
decreases for $\alpha=0$ and, as evident from this tendency,
crosses the separating line at a certain $N$. 
The cases with no amplification even yield the ratios that are considerably
below $1$ although the number of involved particles is reduced by ejection.
This implies that when the amplification of triaxiality seeded by
density fluctuations is masked by the pressure force, the infall of
particles produces more spherical figure than the initial state.
With $\alpha=0.25$, we find that $\overline{\iota_{80}}/\iota_{80,0}$
is, on average, higher than that in the previous case. The points vary
in different way as they
rather fluctuate and drop below the line when $N$ is equal to $32,000$.
Unexpectedly, the points for $N=64,000$ and $128,000$ with $b_{0}=0.3$ 
are located at the upper area despite that the point for $N=32,000$ 
is below the line. We notice remarkably large error bar for $N=64,000$
that spans both regions. This indicates that the obtained
$\overline{\iota_{80}}/\iota_{80,0}$ from two different
realizations are very scattered, with one above and one below
the line separating two regions. This is the outcome from
the finite-$N$ fluctuations where the designation
of $b_{0}$ and $N$ alone cannot determine the final
triaxiality in each individual realization.
Different microscopic arrangements can lead
to the final triaxiality that can be either greater or lower than
the initial one even with $N$ as high as $64,000$.
However, when considering the mean
behavior, the first $N$ where the plot crosses or gets close to
the separating lines, marking the critical number $N_{c}$,
is clearly reduced when $b_{0}$ is higher in both $\alpha$.
This is consistent with
our hypothesis. When $\alpha=0.5$, the transition
is barely seen. This case yields the ratios that are greater
than the two
previous cases and the development of triaxiality is hardly
reduced even though $b_{0}$ is as high as $0.3$: 
only the case with $N=128,000$ lies below the line whereas
the other cases remain far above.

For $\alpha=0$, that $\iota_{80}$ decreases with increasing
$b_{0}$, while still remaining above the separating line,
can also be explained by an extension of our hypothesis beyond
its main purpose for $N_{c}$. According to the
analysis in Section \ref{collapse_eq}, the distinction between
gravity fluctuation- and pressure-dominated core collapse
is made by comparing the values of $r_{J}$ and $\tilde{r}$.
While the core collapse is still dominated by $\tilde{r}$,
the increase of $b_{0}$ simply gives more resisting force,
which is described mathematically by the increase of $r_{J}$
closer to $\tilde{r}$.
This leaves less available space for collapse and amplification
beneath $\tilde{r}$. Therefore, the amplification degree is
reduced, but it is still effective.
When $\alpha >0$, $\overline{\iota_{80}}/\iota_{80,0}$
is higher on average and the same variation with $N$ is
not noticed because the fluctuating pattern is more prominent.
The fact that increasing $\alpha$ tends to yield a more
elevated ratio, albeit with a lower collapse factor
(see Fig. \ref{fig_collapse_b0}), can be
related to the so-called parametric resonance.
This effect is summarized in \citet{levin_et_al_2014_pr}.
In principle, a non-zero $\alpha$ triggers non-simultaneous
collapse: the inner component falls earlier, forming a
dense oscillating core in the process. The resulting oscillating
field is then able to resonate with the late infall and the expansion
of the outer mass shell. This magnifies the extent of the
axis contrast of structure as an additional factor to the 
LMS instability. With higher $\alpha$, the
spread of free-fall times from inside to outside mass shell
is widened. So, there should be more resonances possible
in the violent relaxation, resulting in higher flattening.
In other words, this process of morphological evolution
incorporates another $\alpha$-dependent mechanism, in the form
of a boost, apart from the essential $N$-dependent
LMS instability. This explains why, for $\alpha >0$,
the same measured parameter fluctuates above the threshold. 
It is the remnant of the $\alpha$-dependent virialization.
Note that the parametric resonance is also believed to
account for the formation of core-halo structure,
which is found to be the generic form of the
quasi-stationary state achieved from violent relaxation.

\begin{figure}
  \begin{center}
    \begin{tabular}{c}
      \includegraphics[width=8.0cm]{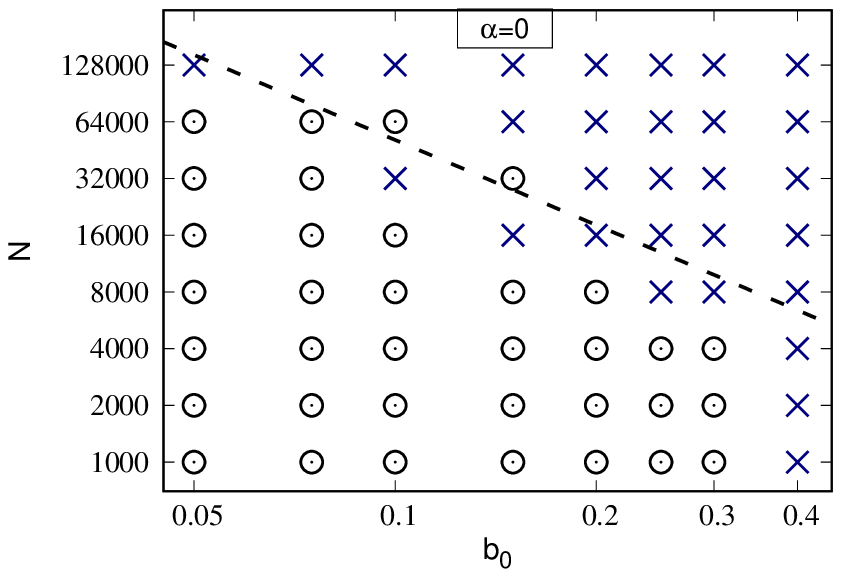} \\
      \includegraphics[width=8.0cm]{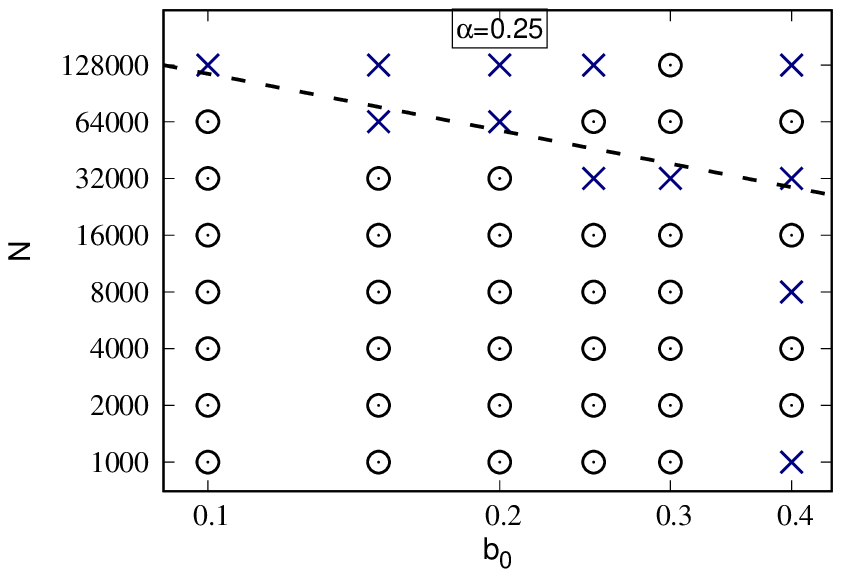}
    \end{tabular}
  \end{center}
  \caption{Summary of $\overline{\iota_{80}}/\iota_{80,0}$ in
    $(b_{0},N)$ space for $\alpha=0$ (top panel) and
    $0.25$ (bottom panel). The cases for any $b_{0}$ and $N$ with
    the ratios above (and below) $1.05$ are depicted in circle
    (and cross). Dash lines are the best-fitting
    theoretical predictions corresponding to $b_{0}^{-3/2}$ and
    $b_{0}^{-11/7}$ for $\alpha=0$ and $0.25$, respectively,
    with $N_{c}$ defined as the lowest $N$ where the ratio turns
    below the threshold. The fitting excludes the result from
    $b_{0}=0.4$ in both cases.} \label{fig_b0_n_space}
\end{figure}

The effectiveness of symmetry breaking is summarized
in $(b_{0},N)$ diagrams in Fig. \ref{fig_b0_n_space} where the
cases with $\overline{\iota_{80}}/\iota_{80,0}$ above and below
$1.05$ are depicted in different ways. By defining $N_{c}$ as
the lowest $N$ where the ratio falls below $1.05$, we provide
the best-fitting lines of $b_{0}^{-3/2}$ and $b_{0}^{-11/7}$
functions with $N_{c}$ for $\alpha =0$ and $0.25$, respectively.
This tests the validity of the hypothesis we gave in
Section \ref{collapse_alphanonzero} and \ref{collapse_alphazero}.
Note that the fitting is performed from the lowest
depicted $b_{0}$ to $b_{0}=0.3$. The case with $\alpha=0.5$,
for which strong symmetry breaking occurs in almost all cases,
is not shown because of insufficient statistics.
The plot for $\alpha=0$ exhibits a clear partition by the
best-fitting line between upper and lower areas, indicating ineffective
and effective amplifications of triaxiality, respectively.
We notice a few anomalies, one for each of $b_{0}=0.1$ and $0.15$,
where $\overline{\iota_{80}}/\iota_{80,0}$ remains higher
than the threshold just above $N_{c}$. Because those cases are
situated near the transition line, different statistical fluctuations
may lead to very different final triaxiality in each realization.
Thus, the ensemble average may have a high dispersion comparing to
the cases that are further away from the line.
The case of $b_{0}=0.4$ is notable and requires further explanation.
The amplification is weak for $N=1,000$, which is far
below the prediction line. This implies over-stability. 
We speculate that when $b_{0}$ is high enough,
the inner pressure force that diverges more rapidly than
the gravitational force becomes so strong that it expels away
the particles in the opposite direction to the infall.
It then attenuates the strength of the triaxial field and leads to
an ineffective symmetry breaking even if $N$ is as low as $1,000$.

For $\alpha=0.25$, we find that $N_{c}$ at each $b_{0}$ is higher
than its $\alpha=0$ counterpart and
it still fits well with the predicted separating
line until $b_{0}=0.3$. Comparing it with the case of $\alpha =0$,
we find a few more anomalies with $b_{0}=0.25$ and $0.3$
where the amplification degree remains high well away from $N_{c}$.
The result for $b_{0}=0.4$ gives rise to more complications since,
at first, it implies the same over-stability because of weak
amplification that is already present with $N=1,000$. 
It then shows a switch between two regimes with 
$N$ up to $N=128,000$. Following from the discussion
on Fig. \ref{fig_ratio_iota_alpha}, these complexities can be
attributed to the combined effect when $\alpha>0$. In the situations
where structural deformation is triggered, as described by our
hypothesis, the violent relaxation of sequential collapse tends
to produce higher triaxiality than that which is produced by
simultaneous collapse given the same amount of initial $\iota_{80}$.
Consequently, the statistical average of the relative $\iota_{80}$,
calculated from the populations of strong and weak amplifications,
has higher possibility to cross the threshold value.
However, this does not contradict our hypothesis
since Fig. \ref{fig_b0_n_space} suggests that, with $\alpha=0.25$,
it simply requires higher $N_{c}$ to meet the chosen criterion and
those numbers still agree with our prediction. 
This proves that the proposed estimate of $N_{c}$ 
can still be valid even if another effect from mild $\alpha$
is incorporated in the evolution.
The situation for $\alpha=0.5$ is that the suppression of
symmetry breaking is captured only when $N=128,000$ with $b_{0}=0.2$
and $0.3$, also excluding $b_{0}=0.4$ due to the same over-stability
referred to above.
The fact that it takes place at an extremity in the 
simulated parameter space leads us to suspect that $N_{c}$ might be
higher for this employed range of $b_{0}$. In other words,
due to a stronger amplification process for $\alpha =0.5$, we possibly
need to further suppress the initial seed by increasing $N$.

As a final remark in this part, we note the presence of a large
dispersion even though $N$ is as high as $64,000$ or more
(see, e.g., Fig. \ref{fig_ratio_iota_alpha}). 
This is understandable since the particles involved in
determining the fate of the final triaxiality are those around
$r_{J}$ or $\tilde{r}$, which account for less than $1 \%$ 
of $N$ in a typical range of $b_{0}$. This fact has not been pointed out
in many reports in the literatures that explored a simulation space
comparable to our high-$N$ range \citep{theis+spurzem_1999,
  roy+perez_2004, boily+athanassoula_2006, barnes+lanzel+williams_2009}.

\subsection{$N$-dependence of other parameters}
\label{n_dependence_collapse}

\begin{figure}
  \begin{center}
    \includegraphics[width=8.5cm]{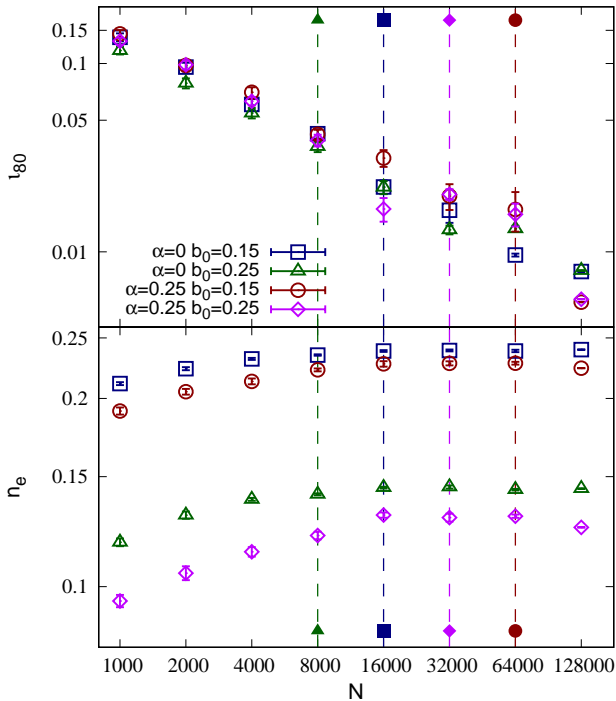}
  \end{center}
  \caption{Ensemble-averaged $\iota_{80}$ (top panel) and $n_{e}$
    (bottom panel) at $9.52 \ t_{d}$ as functions of $N$ for different
    indicated $\alpha$ and $b_{0}$. Vertical dashed lines with 
      filled points designate $N_{c}$ of the plotted cases in empty points
      with the same shapes. Size of error bars is estimated by the
    standard deviation of the mean.} \label{fig_iota_ne_nc}
\end{figure}

In the previous section, we observed the segregation
between strong and weak triaxiality that occurs as consequence of different
dominant factors in the core collapse. These two regimes can be
discriminated efficiently by $\overline{\iota_{80}}/\iota_{80,0}$ and
the transition points behave in accordance with the prediction.
In this section, we consider some additional widely-employed
parameters relating to the collapse and the configuration and examine
whether their variations with $N$ exhibit the same transition
that we reported.
In Fig. \ref{fig_iota_ne_nc} we show the ensemble-averaged
$\iota_{80}$ and $n_{e}$ at $9.52 \ t_{d}$ as functions of $N$ with
various $\alpha$ and $b_{0}$. The vertical dashed lines 
correspond to $N_{c}$ for each case. We consider first
the variation of $\iota_{80}$. The plot exhibits clearly the
decay with $N$. When we inspect the final $\iota_{80}$,
we see that varying $\alpha$ or $b_{0}$ does not strongly
affect the results compared to the variation of $N$.
Thus, it turns out that the influence of both these
parameters is relatively minor below the governance from $N$
in these warm collapse experiments. This is
unlike $\overline{\iota_{80}}/\iota_{80,0}$
(see Fig. \ref{fig_ratio_iota_alpha}), with which the effect
from increasing $\alpha$ and $b_{0}$ can be differentiated.
However, the situation is different for $b_{0}=0$ as
it is found that the influence of $\alpha$ becomes prominent 
as reported by \citet{sylos_labini+benhaiem+joyce_2015}.
From their study, the flattening of relaxed structures
can be enhanced many times by increasing $\alpha$.

While the $N$-dependence of final $\iota_{80}$ is evident
in all plots, its variation does not display the transition
in coherence with the vertical line. The reason why
the transition cannot be spotted from direct examination on
$\iota_{80}$ can be explained as follows. According to the
analysis of \citet*{lin+mestel+shu_1965}, the evolutionary
track of eccentricity depends strongly on the initial value.
It is thus plausible that the
output triaxiality also depends on the initial
strength of seed. In our set-up, the initial $\iota_{80}$ is
governed by the Poissonian noise that can be scaled by $N^{-1/2}$.
Therefore, a strong $N$-dependence of final $\iota_{80}$ can be
understood since it involves strong $N$-dependent effects
from the initial seed to its path of evolution.
This explains why the measurement by
$\overline{\iota_{80}}/\iota_{80,0}$ is more effective:
it weights out the strength of initial fluctuations from
evaluation.

For $n_{e}$, the fraction increases as we increase $N$ before it
apparently attains the terminal value at $N\sim 16,000$.
Before we consider how consistent it is with $N_{c}$, we should
explain this particular pattern of $n_{e}$.
First, we recall the analysis of cold collapse
by \citet{joyce+marcos+sylos_labini_2009}, which demonstrates
that the collapse of a uniform sphere is more condensed as
$N$ increases due to smaller density fluctuations. This makes
$n_{e}$ increase accordingly because of the stronger kinetic kick
by $N$-dependent central condensation.
This statement is presumably applicable to our low-$\alpha$
collapse with low $N$ as we observe an analogous variation
of $n_{e}$. In this range, the minimal size dictated by
density fluctuations is not particularly small so the system can
attain this size and re-expand before the bouncing
force from the initial pressure can take any action. In other words,
the contraction and ejection in this regime are dominantly governed
by density fluctuations that leads to the $N$-dependent consequence.
When $N$ is high enough, reduced density fluctuations attempt
to produce a more condensed collapse before it is eventually
prevented by the pressure that limits the collapse to
progress further. Thus, in this scenario, 
the pressure replaces the density fluctuations as the
major actor in controlling the mass concentration and the
amount of ejection.
The fact that the pressure is independent of $N$ and that its
fluctuations decay more rapidly with $N$
leads us to deduce that the achieved
maximum contraction and resulting ejection are less
relying on $N$ than is its low-$N$ counterpart. This explanation
also applies to the plots of $n_{e}$ at high $N$.

Although the proposed mechanism to describe the plot
of $n_{e}$ implies interplay between gravity and
pressure, their transition points do not coincide with the lines
of principal transition since these points appear to
be displaced more weakly along the $N$-axis. The weak $N$-dependence
of the transition points in $n_{e}$ might be because the determination
of them does not fully include the fluctuations of the two
counter-acting forces, both of which depend strongly on $N$. 
It is unlike the shape deformation instability that takes
the fluctuations of all forces into consideration.

In summary, the demonstrated plots of some standard parameters
emphasize once again the usefulness of
$\overline{\iota_{80}}/\iota_{80,0}$ in distinguishing between
different regimes to test our hypothesis. The transition
is completely undetectable in some parameters such as $\iota_{80}$. 
Although other parameters, e.g. $n_{e}$, appear to exhibit
the related transition, these points do not reflect any proximity
to our proposed transition.

\section{Conclusion} \label{conclusion}

In this work, we numerically investigate the spherical collapse
of gravitationally unstable $N$-body systems that cover
a considerable range over the power-law index of density $\alpha$,
the particle number $N$ and the initial virial ratio $b_{0}$.
Our main objective is to resolve whether the collapse exhibits different
regime of shape evolution as we vary the system parameters.
In parallel, we develop the collapse model
used in the pioneering Lin-Mestel-Shu theory \citep*{lin+mestel+shu_1965}
by introducing the velocity dispersion to the initial state.
For an unstable sphere of particles, 
we first evaluate the profiles of the gravitational and pressure forces
and their fluctuations, involving the Poissonian noise in
both spaces. By finding the balances of these
forces, two effective radii are obtained. The first one fixes
the size of the non-collapsing region similar to the conventional
Jeans' length while the second one indicates at which scale the
density fluctuations are dominant over the pressure fluctuations.
We regard the latter radius as the size of the triaxial seed
from density fluctuations. Depending on which radius is dominant,
two different scenarios of shape evolutions are proposed.
When the latter radius is dominant, 
the collapsing core develops further the triaxiality from the
Poissonian density fluctuations and leads to the amplification
of entire structure accordingly. Otherwise, the amplification of core
triaxiality is halted and the collapsing system remains weakly
amplified because it lacks a strong triaxial field.
The point where the two radii are equal marks the
transition, which is parametrized by the critical $N$ above which
the amplification turns ineffective.
Although our analysis is specific to simple density profile and velocity
distribution, we believe that this methodology can also be
applicable to other choices of initial states as long as the Poissonian
fluctuations are involved. Different distribution functions simply
yield different force profiles and values of the relevant parameters.

Considering the numerical results, the tracing of the
flattening of different fractions of bound mass justifies the
inside-out shape evolution. To inspect more fully our hypothesis,
we measure the ratio of the main flattening parameter between the virialized
and initial states (or $\overline{\iota_{80}}/\iota_{80,0}$) as the main tool.
This measurement proves
effective as it weights out strong $N$-dependence from initial
numerical value fixed by the initial density fluctuations.
In doing this, many crucial results are unveiled.
First, as supposed to be the utility of it, the segregation
between effective and ineffective symmetry breaking by the
critical particle number is definitely identified.
The variation of critical $N$ as function
of $b_{0}$ agrees with our prediction in the range of 
lowest performed $b_{0}$ to $0.3$ provided that
$\alpha \leq 0.25$. As $b_{0}$ reaches $0.4$, the flattening
process appears to be inactivated in almost all cases.
We speculate that this is because of excessive
pressure force that expels away the particles in the core.
About the influence of $\alpha$, the increase of it from $0$
to $0.25$ not only produces higher flattening, as suggested by
past literatures, but it also raises the critical $N$
for the same $b_{0}$. This is possibly due to another 
$\alpha$-dependent amplifying process from non-simultaneous
infall in addition to that fuelled by the LMS instability.
The fact that our prediction is still quantitatively
valid despite the involvement from mild $\alpha$ assures
that the hypothesis of the interplay between two competing
forces at two orders in the core is applicable.
The situation for $\alpha=0.5$ is more
complicated but it is however comprehensible by the same
speculation. In this case, the involvement from $\alpha$
plays more significant role to further elevate the ratio.
Even faint triaxiality at start
could end up highly flattened by the combined
procedure. As consequence, almost all cases remain far from
initial flattening and only few cases with highest $N$ and high
$b_{0}$ are subdued. This suggests that the critical $N$ is higher.

From the conclusion, it appears that the
gravitational collapse is more complicated
than the proposed description that is on the basis of the
self-consistent LMS framework alone. Other known astrophysical
mechanisms such as the parametric resonance or the radial
orbit instability might also come into play 
even if the initial density profile slightly deviates from
homogeneity. The combined process leads to another degree
of amplification. This makes us uncertain about the validity
of our hypothesis beyond mildly inhomogeneous density profile
since the non-LMS driving process is more prominent.

\section{Discussion} \label{discussion}

In this section, we discuss the related studies in the past.
From analytical point of view, we note an analogous work
of \citet{boily+athanassoula+kroupa_2002} where they also
analyzed the cold collapse to determine 
the critical particle number. Their number indicates
other transition where, exceeding this number, the Poissonian
fluctuations cease to have influence on a mildly ellipsoidal collapse.
Consequently, the evolution strictly follows the LMS pattern.
This work has the same conception such that, above the
critical particle number, the influence of Poissonian fluctuations
on the violent relaxation is limited.
However, without pre-adjusted ellipticity, a cold spherical collapse is
always unstable according to our hypothesis since
there is no pressure. Considering then the numerical
results, our work demonstrates the role of $N$
in quite the same way as shown by \citet{benhaiem_et_al_2018}.
This group carries out the spherical collapse experiments
with $\alpha=1$ and,
as first expectation, obtains far greater $\iota_{80}$
compared to our $\alpha=0.5$ simulations. Similarly, they exemplify
the suppression of the onset of the radial orbit instability
following the collapse by increasing $N$. While the driving
mechanism is different, the fact that reducing the amplitude
of initial fluctuations by increasing $N$ is able to turn off
the key process that breaks strongly the symmetry relates closely
to our scope. However, more quantitative evaluation
of the transition point has been omitted there.
About the attempt to retrieve the critical values,
\citet{min+choi_1989} have investigated the collapse of a
sphere with uniform density and proposed an upper limit with
$b_{0}=0.3$, below which the symmetry breaking is efficient.
Next, \citet{boily+athanassoula_2006} have performed a similar
investigation for the cases with non-uniform density profiles
and reported the limiting $b_{0}=0.4$. This latter value coincides with
a part of our summary but this suppression is caused by
an excessive kinetic energy in the core which is out of the main scope.

Before we continue the discussion, we should make the relation
between the parameters clear. 
Although their employed parameter to pinpoint the transition is
different from ours, the diagrams in Fig. \ref{fig_b0_n_space} 
suggest that these two parameters are interchangeable:
we can draw either the critical $N$ at the given $b_{0}$ or the
critical $b_{0}$ at the given $N$. According to our finding,
it appears that some crucial points are missing in those works.
First, the practice to examine the final flattening parameter
without considering the initial value and then to specify
an arbitrarily small value for a transition might not be sufficient to
give an accountable critical point. Note that there are also
the temporal and ensemble fluctuations in the relaxed states.
This causes, on the one hand,
the improper identification of the critical value or, on the other hand,
the unawareness of the transition while varying $b_{0}$
\citep{van_albada_1982, mcglynn_1984, roy+perez_2004}.
The next point is that a certain critical $b_{0}$ is actually not
applicable to all systems since we find that it changes significantly
from one initial state to another. For example, given the same
initial density profile, the critical $b_{0}$ significantly shifts as
we simply change $N$. We would additionally mention that in some cases,
e.g. the power-law density profile with high exponent,
the critical number might be beyond the executable $N$, if evaluated
in our way. Our results with $\alpha=0.5$ already give an
example of that claim.

\section*{Acknowledgements} \label{acknow}
This research is supported by the Thailand Research Fund (TRF) and
Rajamangala University of Technology Suvarnabhumi via the
Grant for New Researcher with contract number TRG5880036
under the mentorship of Khamphee Karwan. Numerical simulations
are facilitated by HPC resources of Chalawan cluster of the
National Astronomical Research Institute of Thailand (NARIT).
The comments from the anonymous reviewer and Michael F. Smith are
also gratefully acknowledged.

\section*{Data availability} \label{acknow}
The data used in this article can be shared on reasonable
request to the corresponding author.

\bibliographystyle{mnras}

\label{lastpage}

\end{document}